\documentclass{emulateapj}
\RequirePackage{lineno}
\usepackage{amssymb}
\usepackage{amsmath}
\usepackage{mathrsfs}
\usepackage{natbib}
\usepackage[colorlinks, linkcolor=blue, urlcolor=blue, citecolor=blue]{hyperref}
\usepackage{array}
\usepackage{float}
\usepackage{graphicx}
\usepackage{subfigure}
\usepackage{color}
\usepackage[all]{hypcap}
\usepackage[toc,page]{appendix}
\usepackage{diagbox}
\usepackage{booktabs}
\usepackage{multirow}

\def\beq{\begin{equation}}
\def\enq{\end{equation}}
\def\bea{\begin{eqnarray}}
\def\ena{\end{eqnarray}}

\begin{document}

\title{On the properties of a newborn magnetar powering the X-ray transient CDF-S XT2}
\author{Di Xiao\altaffilmark{1,2}, Bin-Bin Zhang\altaffilmark{1,2}, and Zi-Gao Dai\altaffilmark{1,2}}
\affil{\altaffilmark{1}School of Astronomy and Space Science, Nanjing University, Nanjing 210093, China; dxiao@nju.edu.cn; bbzhang@nju.edu.cn; dzg@nju.edu.cn}
\affil{\altaffilmark{2}Key Laboratory of Modern Astronomy and Astrophysics (Nanjing University), Ministry of Education, China}

\begin{abstract}
Very recently \citet{XueYQ2019} reported an important detection of the X-ray transient, CDF-S XT2, whose light curve is analogous to X-ray plateau features of gamma-ray burst afterglows. They suggested that this transient is powered by a remnant stable magnetar from a binary neutron star merger since several pieces of evidence (host galaxy, location, and event rate) all point toward such an assumption. In this paper, we revisit this scenario and confirm that this X-ray emission can be well explained by the internal gradual magnetic dissipation process in an ultra-relativistic wind of the newborn magnetar. We show that both the light curve and spectral evolution of CDF-S XT2 can be well fitted by such a model. Furthermore, we can probe some key properties of the central magnetar, such as its initial spin period, surface magnetic field strength and wind saturation Lorentz factor.
\end{abstract}

\keywords{stars: neutron -- radiation mechanisms: general -- X-rays: individual}

\section{Introduction}
Ever since the discovery of the first binary neutron star (NS) merger event GW170817 \citep{Abbott2017a}, there is remarkable progress on the study of gravitational waves and multi-wavelength counterparts. A few important issues have been explored through the rich multi-messenger observational data of GW170817 \citep{Abbott2017b}, such as the jet structure of gamma-ray burst (GRB), energy source of kilonova, the equation of state (EOS) of NS and so on. However, one key problem remains unsolved, which is the identification of the remnant of the binary NS merger. There is no signal found from the search for post-merger gravitational waves from the remnant \citep{Abbott2017c}. Therefore we could not identify the remnant directly. A few pieces of indirect evidence of stable supermassive NS formation have been proposed since it seems that an energy injection from a newborn NS is needed to fit the multi-wavelength afterglow \citep{GengJJ2018}, kilonova emission \citep{YuYW2018, AiSK2018, LiSZ2018}, and a late-time X-ray flare \citep{Piro2019}. However, the black hole (BH) central engine could not be completely ruled out. Thus, the remnant of GW170817 remains a mystery due to a lack of ``smoking gun" evidence.

The electromagnetic (EM) signals differ in many aspects whether a BH or stable NS is formed from binary NS merger, as have been discussed in \citet{Metzger2012} and \citet{GaoH2013}. If a stable NS is formed, a spin-down energy injection is naturally expected, and the EM signals are generally brighter than those in the situation of BH central engine. Firstly, there could be plateaus or flares in the X-ray afterglow light curves of associated short GRBs \citep{DaiZG1998a, DaiZG1998b, DaiZG2006, ZhangB2006}. Secondly, the associated kilonovae can reach a much higher luminosity due to energy injection, which was named as ``Mergernovae" \citep{YuYW2013}. Thirdly, the sub-relativistic ejecta can be accelerated to relativistic speed. Hence, the emission from ejecta-interstellar medium interactions could be much brighter \citep{GaoH2013, WuXF2014}. Moreover, a unique counterpart of X-ray emission is suggested from the internal dissipation in an ultra-relativistic quasi-isotropic wind of the newborn NS \citep{ZhangB2013, Metzger2014}, which is not expected for a BH central engine. If the observer is off-axis from the short GRB and there is little ejecta matter in the line of sight (as shown in Figure 1 of \citet{GaoH2013}), this X-ray emission is the only EM signal that can be observed from a binary NS merger, whose different possible light curves have been modeled in \citet{SunH2017}. In this Letter, we propose that the newly-discovered X-ray transient CDF-S XT2 is exactly this kind of signal.

The light curve of CDF-S XT2 is analogous to the X-ray plateau feature of a GRB afterglow \citep{XueYQ2019}, which is thought to be the signature of a long-lasting energy injection from a newborn magnetar \citep{DaiZG1998a, DaiZG1998b, ZhangB2001, ZhangB2006, YuYW2009, YuYW2010, DallOsso2011, Stratta2018}. However, the absence of prompt gamma-ray emission suggests that we are off the axis of a GRB. The emission of CDF-S XT2 should have ``internal" origin since it is not seen at optical or radio band. High-energy emission is naturally expected as the magnetic energy of a quasi-isotropic magnetar wind gradually dissipates via magnetic reconnection \citep{Spruit2001, Drenkhahn2002b, Giannios2005, Metzger2011, Beniamini2014, Beniamini2017b, XiaoD2017, XiaoD2018}. As we have proposed in \citet{XiaoD2019}, the internal X-ray plateaus of GRBs can be well explained within this scenario. This model applies perfectly to CDF-S XT2 not only from the light curve but also from its spectral evolution. Observationally, a transition of X-ray photon index from $1.57_{-0.50}^{+0.55}$ before 2000\,s to $2.53_{-0.64}^{+0.74}$ after 2000\,s is reported \citep{XueYQ2019}, which matches the model prediction of spectral evolution from $F_{\nu}\propto \nu^{-0.5}$ to $F_{\nu}\propto \nu^{-(p-1)/2}$ well \citep{XiaoD2019}. Comparing with the observation, we can obtain the power-law index of the electrons accelerated by magnetic reconnection, $p=4.06_{-1.28}^{+1.48}$. Latest Particle-in-Cell simulations suggest that the electron power law index accelerated by magnetic reconnection is $1\leq p\leq2$  if the magnetization $\sigma\gg 1$, and $2\leq p\leq4$ if $1\leq\sigma\leq 10$ \citep[e.g.,][]{Sironi2014, GuoF2015, GuoF2016}. In the gradual magnetic dissipation model discussed in this work, non-thermal emission is produced from the photospheric radius to the saturation radius, at which $\sigma\sim10$  and $\sigma=1$ respectively \citep{Beniamini2017b, XiaoD2017}. Therefore, the above electron power-law index, $p=4.06_{-1.28}^{+1.48}$, considering the large error bars, is marginally consistent with the simulation results.

This paper is organized as follows. We present the method of light curve fitting and the application to CDF-S XT2 in section \ref{sec2}. Then in Section \ref{sec3} we constrain the properties of the central magnetar from the fitting results. We finish with discussion and conclusions in Section \ref{sec4}.

\section{Fitting the light curve of CDF-S XT2}
\label{sec2}
A newborn magnetar loses its rotational energy via gravitational-wave and electromagnetic radiation, whose angular velocity evolution can be generalized as follows ,
\beq
\dot{\Omega}=-k\Omega^n,
\label{eq2}
\enq
where $\Omega=\Omega(t)=2\pi/P(t)$ is the spin angular velocity, and $k$ and $n$ represent a constant of proportionality and the braking index of magnetar respectively. When several different torques are acting at the same time, Eq.(\ref{eq2}) can be regarded as an ``effective torque" equation and $n$ as an effective braking index, as done recently by several works \citep[e.g.,][]{Lasky2017,LvHJ2019}. The solution of Eq.(\ref{eq2}) is
\beq
\Omega(t)=\Omega_0\left(1+\frac{t}{\tau}\right)^{\frac{1}{1-n}}
\label{eq3}
\enq
where $\Omega_0$ is the initial angular velocity and $\tau\equiv\Omega_0^{1-n}/[(n-1)k]$ is the spin-down timescale. The injected energy comes from the magnetic dipole torque whose luminosity is $L_{\rm EM}=B^2R^6\Omega^4/6c^3$. Therefore, the observed X-ray flux is
\bea
F_{X,\,\rm obs}&=&(1+z)\eta_X L_{\rm EM}/4\pi D_L^2 \nonumber\\
&=&\frac{1+z}{4\pi D_L^2}\eta_X L_0\left(1+\frac{t+t_0}{\tau}\right)^{\frac{4}{1-n}} ,
\label{eq4}
\ena
where $t_0$ accounts for the possible delay between magnetar formation and its X-ray emission \citep{Metzger2011}, $L_0\equiv B^2R^6\Omega_0^4/6c^3=1.0\times10^{49}B_{15}^{2}R_6^6P_{0,-3}^{-4}\,\rm erg\,s^{-1}$, $z$ is redshift and $D_L$ is the corresponding luminosity distance. The X-ray radiation efficiency $\eta_X$ depends strongly on the injected luminosity $L_{\rm EM}$ \citep{XiaoD2019}.

To obtain the relation $\eta_X=\eta_X(L_{\rm EM})$, we should start from the radiation mechanism of this high-energy emission. A newborn rapidly-rotating magnetar can produce a Poynting-flux-dominated wind \citep{Aharonian2012}, the magnetic field lines of which could be in a ``striped wind" configuration \citep{Coroniti1990,Spruit2001}. The high-energy emission from the internal gradual magnetic dissipation process in the wind has been discussed in detail \citep{Beniamini2017b, XiaoD2017, XiaoD2018}. Since the initial magnetization $\sigma_0$ of the wind is unknown, we consider five cases of different wind saturation Lorentz factor $\Gamma_{\rm sat}$, which is equivalent to $\sigma_0$ since $\Gamma_{\rm sat}=\sigma_0^{3/2}$ \citep{Beniamini2017b}. The values $\Gamma_{\rm sat}=10^2,\,10^{2.5},\,10^3,\,10^4,\,10^5$ are adopted following the calculation in \citet{XiaoD2019}. Since $\eta_X$ is dependent on the observational properties such as  the energy range of the instrument and redshift of the source, it is not easy to derive an analytical relation. Instead, we can carry out an empirical polynomial fitting to obtain the X-ray efficiency $\eta_X$ as a function of injected electromagnetic luminosity $L_{\rm EM}$, which are
\bea
\log\eta_X = -0.033(\log L_{\rm EM})^2 + 2.91\log L_{\rm EM}-65.66, \nonumber\\
\log\eta_X = -0.064(\log L_{\rm EM})^2 + 6.09\log L_{\rm EM}-144.95, \nonumber\\
\log\eta_X = -0.039(\log L_{\rm EM})^2 + 3.87\log L_{\rm EM}-98.71, \nonumber\\
\log\eta_X = -0.006(\log L_{\rm EM})^2 - 0.27\log L_{\rm EM}-4.61, \nonumber\\
\log\eta_X = -0.015(\log L_{\rm EM})^2 - 1.15\log L_{\rm EM}+14.82, \nonumber\\
\label{eta}
\ena
for $\Gamma_{\rm sat}=10^2,\,10^{2.5},\,10^3,\,10^4,\,10^5$ respectively. The dependence of X-ray efficiency on the injected luminosity will influence the X-ray temporal decay index after plateau phase.

\begin{figure}
\label{fig1}
\begin{center}
\includegraphics[width=0.45\textwidth]{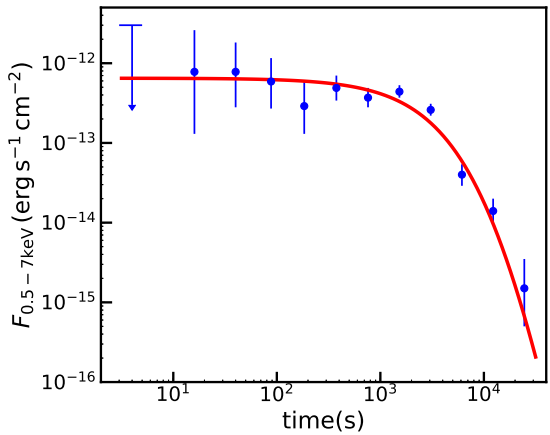}
\caption{The fitting (red line) to X-ray data (blue points) of CDF-S XT2 for $\Gamma_{\rm sat}=10^4$ case.}
\end{center}
\end{figure}

\begin{figure*}
\label{fig2} \centering\includegraphics[angle=0,height=7.5in]{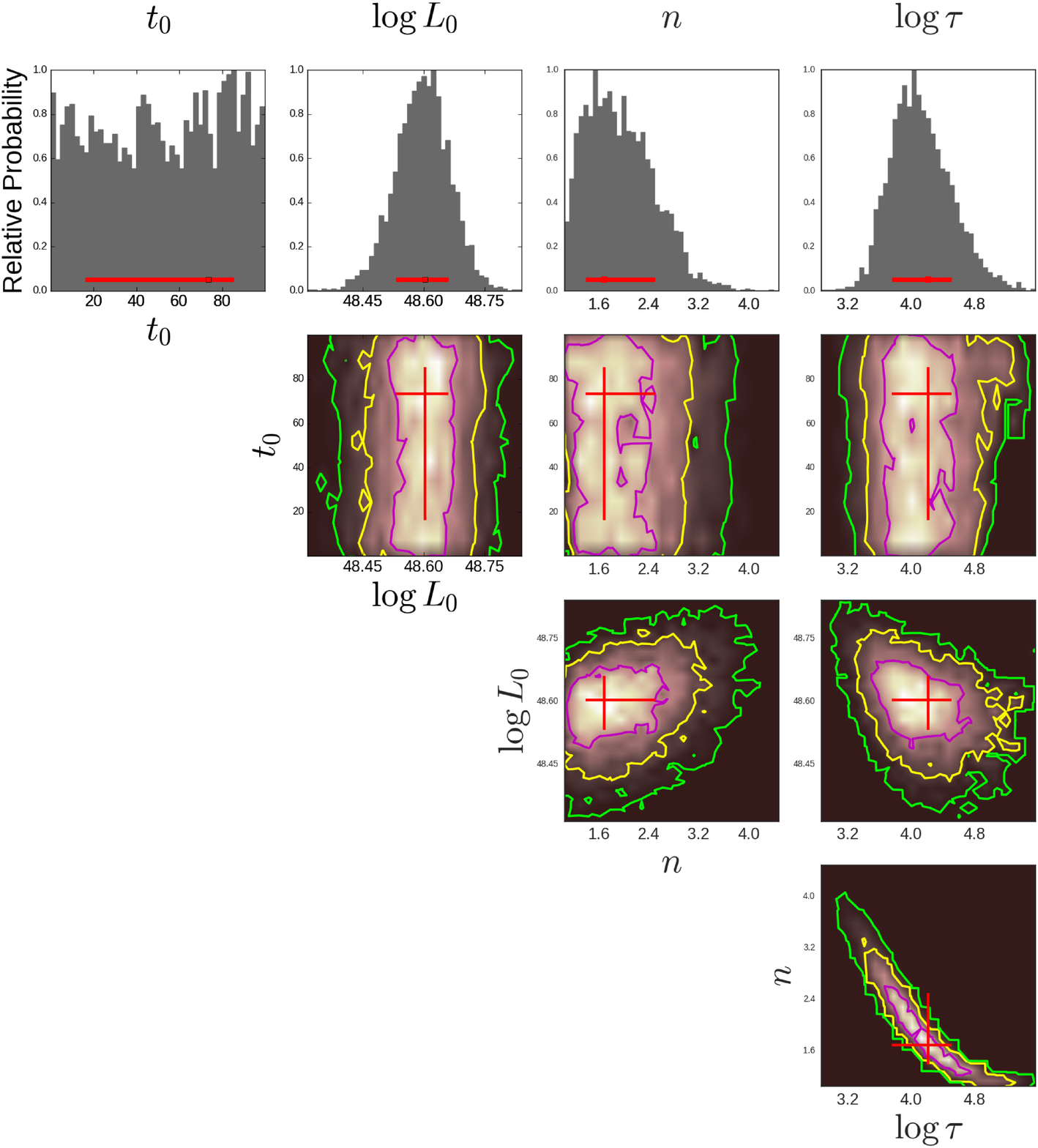} \ \
\caption{Parameter constraints of light curve fitting for $\Gamma_{\rm sat}=10^4$ case. Histograms and contours illustrate the likelihood map. Red crosses show the best-fitting values and
1-sigma error bars.}
\end{figure*}

Taking $(t_0,\,L_0,\,n,\,\tau)$ as parameters, now we can do a Bayesian Monte-Carlo fitting using MCurveFit package \citep{ZhangBB2016}. The best-fitting parameters are shown in Table \ref{table1}. As an example, we show the light curve fitting to the X-ray data of CDF-S XT2 for $\Gamma_{\rm sat}=10^4$ case in Figure \ref{fig1} and the parameter corner for this case is shown in Figure \ref{fig2}.

\begin{table}
\centering
\caption{The best-fitting parameters for five different $\Gamma_{\rm sat}$.}
\begin{tabular}{ccccccccc}
\hline
\hline
\multirow{2}{*}{} & \multicolumn{4}{c}{Best-fitting values} \\
\cline{2-5}& $t_0$ & $\log L_0$ & $n$ & $\log \tau$ \\\hline
$\Gamma_{\rm sat}=10^2$ & $72.72_{-55.26}^{+10.77}$ & $46.14_{-0.11}^{+0.09}$ & $1.59_{-0.20}^{+0.66}$ & $4.12_{-0.48}^{+0.23}$ \\
$\Gamma_{\rm sat}=10^{2.5}$ & $74.53_{-59.64}^{+9.51}$ & $46.24_{-0.08}^{+0.08}$ & $1.76_{-0.30}^{+0.86}$ & $4.13_{-0.48}^{+0.26}$ \\
$\Gamma_{\rm sat}=10^3$ & $61.76_{-44.97}^{+23.07}$ & $46.96_{-0.07}^{+0.07}$ & $1.83_{-0.32}^{+0.84}$ & $4.12_{-0.45}^{+0.26}$ \\
$\Gamma_{\rm sat}=10^4$ & $73.47_{-57.33}^{+12.11}$ & $48.60_{-0.07}^{+0.06}$ & $1.69_{-0.30}^{+0.81}$ & $4.22_{-0.46}^{+0.30}$ \\
$\Gamma_{\rm sat}=10^5$ & $92.33_{-33.97}^{+7.65}$ & $50.21_{-0.07}^{+0.06}$ & $1.72_{-0.30}^{+0.84}$ & $4.22_{-0.47}^{+0.28}$ \\
\hline
\hline
\end{tabular}
\label{table1}
\end{table}

\section{Constraining the stellar properties}
\label{sec3}
\begin{figure}
\label{fig3}
\begin{center}
\includegraphics[width=0.45\textwidth]{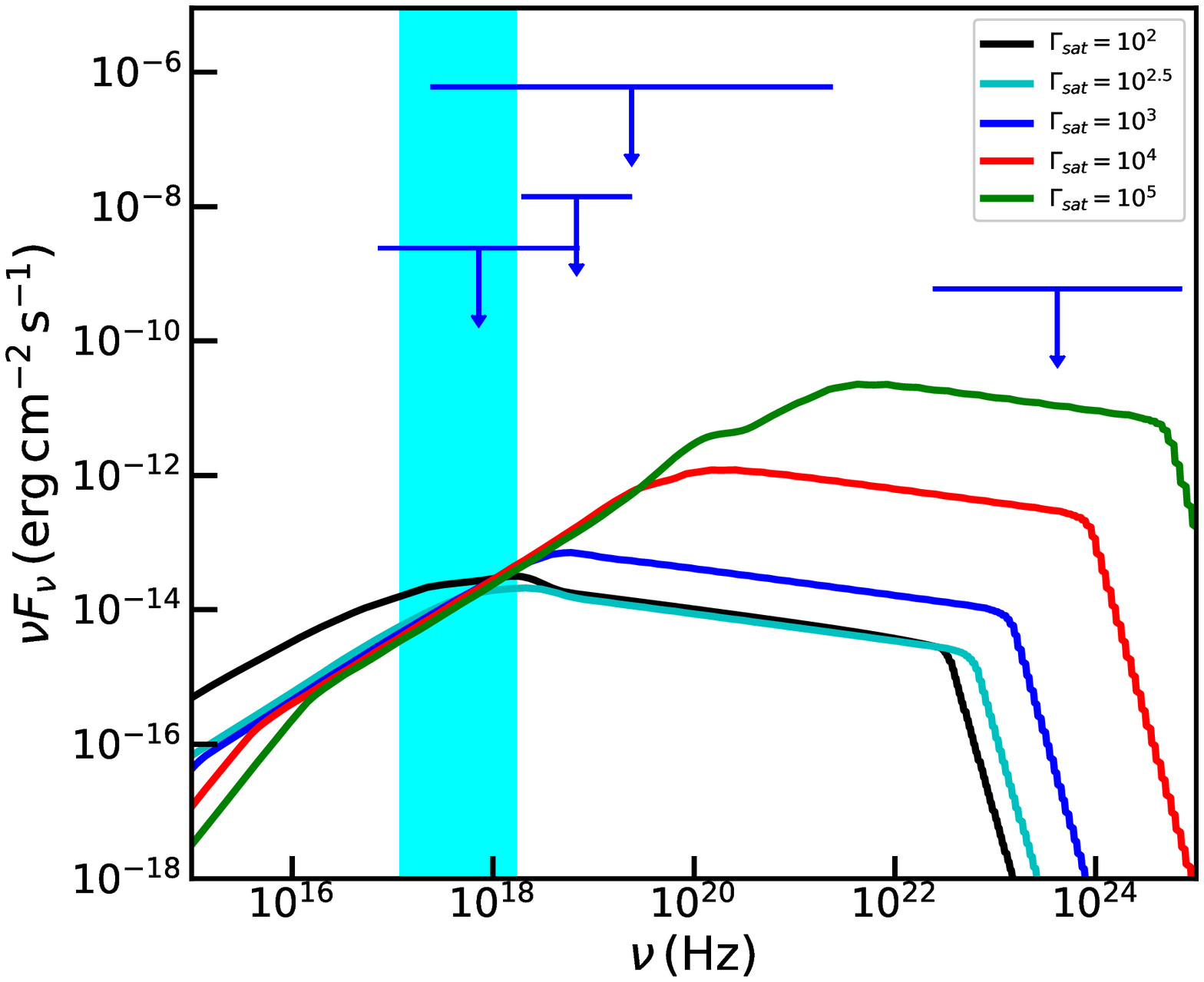}
\caption{The radiation spectrum corresponding to different values of $\Gamma_{\rm sat}$. The value of $\Gamma_{\rm sat}$ uniquely determine the NS spin period and magnetic dipole field, as shown in Table \ref{table2}. Different lines represent different $\Gamma_{\rm sat}$ indicated in the upper-right corner. Four blue upper limits are obtained from high-energy observation. Cyan region represents the observational frequency range of {\it Chandra}.}
\end{center}
\end{figure}

With the best-fitting parameters we can probe the central magnetar in several aspects. Since the deduced braking index $n<3$, besides the magnetic dipole torque, another braking mechanism should play an important role. For instance, fall-back accretion onto the magnetar could lead to $n<3$ \citep{Metzger2018}. This braking index is not surprising as a systematic study of a large sample of GRBs (long and short) with X-ray plateaus also suggested $n$ significantly smaller than 3 \citep{Stratta2018}. Anyway, the deduced timescale $\tau$ in Table \ref{table1} should not be longer than the spin-down timescale purely by magnetic dipole torque $\tau_{\rm EM}$, which means that $\tau\lesssim\tau_{\rm EM}=2\times10^3\,{\rm s}\, I_{45}B_{15}^{-2}R_6^{-6}P_{0,-3}^2$. Combining with the definition of $L_0$ below Eq.(\ref{eq4}), if typically $R_6^6\sim1$ and $I_{45}\sim1.9$ is assumed for the remnant supramassive magnetar \citep{Hotokezaka2013, Piro2017}, we can obtain the upper limits of initial spin period $P_0$ and magnetic field strength $B$. The results are shown in Table \ref{table2}. With these values we can calculate the emission from the gradual magnetic dissipation process within the magnetar wind, which is composed of a thermal component and a nonthermal synchrotron component \citep{Beniamini2017b, XiaoD2017}. Here we plot the radiation spectrum in Figure \ref{fig3} and compare with the flux upper limit of high-energy emission from observations. As reported by \citet{XueYQ2019}, the flux upper limits are $f_{1-10^4\,\rm keV}=6.0_{-0.7}^{+0.7}\times10^{-7}\,\rm erg\,cm^{-2}\,s^{-1}$, $f_{0.3-30\,\rm keV}=2.4_{-2.1}^{+5.3}\times10^{-9}\,\rm erg\,cm^{-2}\,s^{-1}$, $f_{8-100\,\rm keV}=1.4_{-0.3}^{+0.3}\times10^{-8}\,\rm erg\,cm^{-2}\,s^{-1}$, $f_{100\,\rm MeV-30\,\rm GeV}=6.0\times10^{-10}\,\rm erg\,cm^{-2}\,s^{-1}$, respectively, which are also indicated in Figure \ref{fig3}. We can see that the constraint from high-energy emission observation is not very tight and all five cases do not violate these limits.

\begin{table}
\centering
\caption{The upper limits of initial spin period and magnetic field strength for five different $\Gamma_{\rm sat}$.}
\begin{tabular}{ccccccccc}
\hline
\hline
& $P_0$ (in ms) & $B$ (in Gauss)\\\hline
$\Gamma_{\rm sat}=10^2$ & $14.35_{-4.50}^{+13.67}$ & $7.68_{-4.06}^{+21.62}\times10^{15}$ \\
$\Gamma_{\rm sat}=10^{2.5}$ & $12.82_{-4.17}^{+11.80}$ & $6.83_{-3.72}^{+18.33}\times10^{15}$ \\
$\Gamma_{\rm sat}=10^3$ & $5.64_{-1.78}^{+4.58}$ & $3.04_{-1.61}^{+6.92}\times10^{15}$ \\
$\Gamma_{\rm sat}=10^4$ & $0.76_{-0.25}^{+0.64}$ & $3.63_{-2.02}^{+8.78}\times10^{14}$ \\
$\Gamma_{\rm sat}=10^5$ & $0.12_{-0.04}^{+0.10}$ & $5.73_{-3.12}^{+14.05}\times10^{13}$ \\
\hline
\hline
\end{tabular}
\label{table2}
\end{table}

Typically, a ``millisecond magnetar" formed by neutron star mergers has an initial spin period around 1 ms \citep{DaiZG1998a, DaiZG1998b}, as confirmed by the latest numerical simulation \citep{Kiuchi2018}. Also, the magnetic field strength generated by either $\alpha-\Omega$ dynamo \citep{Duncan1992} or amplification of initial field through shear instabilities \citep{Balbus1991, Price2006, Zrake2013} are suggested in the range $\sim10^{14}-10^{15}\,\rm G$, which is also consistent with the simulation results \citep{Kiuchi2014}. As we can see in Table \ref{table2}, for $\Gamma_{\rm sat}=10^2,\,10^{2.5}$ cases, the magnetar rotates too slowly and the magnetic field strength is very high. The $\Gamma_{\rm sat}=10^5$ case goes to the other extreme that the spin period is sub-millisecond. These extreme cases are highly unlikely. For the reasons discussed above, the scenario proposed here would work best for $\Gamma_{\rm sat}\sim10^3-10^4$, which happens to be quite plausible given our current understanding of magnetar winds.

\section{Discussion and Conclusions}
\label{sec4}
In this work, we have provided a theoretical model of the radiation mechanism for the newly-discovered X-ray transient CDF-S XT2. This X-ray emission originates from the internal magnetic dissipation within the quasi-isotropic wind of a newborn magnetar. Both its light curve and spectral evolution are well within the expectation of this scenario. At the beginning the observed frequency of {\it Chandra} satisfies $\nu_c< \nu_a< \nu_{\rm obs}< \nu_m$ and then turns into $\nu_a< (\nu_m, \nu_{\rm obs}) < \nu_c$ later. Correspondingly the synchrotron spectrum evolves from $L_\nu \propto \nu^{-0.5}$ to $L_\nu \propto \nu^{-(p-1)/2}$ \citep{XiaoD2019}. This prediction is verified by the observed X-ray photon index of CDF-S XT2 \citep{XueYQ2019}. Also, the deduced electron power-law index is marginally consistent with the simulation results. We obtained the initial EM luminosity, braking index and spin-down timescale by the fitting of the light curve. Furthermore, by comparing with the numerical simulation results of binary NS mergers, the initial spin period, the magnetic field strength of the central magnetar and the wind saturation Lorentz factor can be constrained. Reasonable values of $P_0\sim1\,\rm ms$, $B\sim10^{14}-10^{15}\,\rm G$ and $\Gamma_{\rm sat}\sim10^3-10^4$ can be reached.

This kind of high-energy emission is only expected if the remnant of binary NS merger is a stable supermassive NS and the discovery of CDF-S XT2 provides strong evidence for this. This emission can be seen at a larger observing angle than short GRB prompt emission \citep{GaoH2013}. Therefore, it has a better chance to be observed. This new EM signal from binary NS merger is a unique probe for the remnant NS, and we can use it to study the physics of newborn magnetar. Further, constraining the EOS of NS is also possible.

A general prediction of the model in this work is that there will be gamma-ray emission at the same time of X-ray emission. However, as we can see in Fig \ref{fig3}, the simultaneous gamma-ray flux is below the detection threshold of {\it Swift}-BAT and {\it Fermi}-GBM. Still, it is possible to observe the gamma-rays if a similar event happens at a closer distance in the future. Also, more facilities with better sensitivity (e.g., {\it Insight}-HXMT) could be critical in finding more similar events like CDF-S XT2.

\acknowledgements

This work is supported by the National Key Research and Development Program of China (Grant No. 2017YFA0402600) and the National Natural Science Foundation of China (Grant No. 11573014, 11833003, and 11851305). DX is also supported by the Natural Science Foundation for the Youth of Jiangsu Province (Grant NO. BK20180324). BBZ acknowledge the support from the National Thousand Young Talents program of China.

\end{document}